\documentclass[twoside]{ilcws10}
\usepackage[latin1]{inputenc}
\usepackage[dvips]{graphicx,epsfig,color}
\usepackage{wrapfig,rotating}
\usepackage{amssymb,amsmath,array}

\begin{document}

\title{CALICE Second Generation AHCAL Developments }

\author{Riccardo Fabbri
\vspace{.3cm}\\
on behalf of the CALICE Collaboration\\
FLC, DESY, Notkestrasse 85, 
22607 Hamburg, Germany\\
E-mail: Riccardo.Fabbri@desy.de\\
}

\maketitle

\begin{abstract}
The CALICE Collaboration is developing and commissioning a 
technological prototype of a hadronic sandwich calorimeter
with approximately 2500 scintillating plates, individually read out 
by multi-pixel silicon photomultipliers. 
The new prototype for the AHCAL aims to demonstrate the feasibility 
to build a calorimeter with fully integrated electronics meeting the 
constraints of a real detector for the International Linear Collider. 
The concept of the prototype, as well as the first results obtained
in the on going test-beam campaign at DESY are reported here.
\end{abstract}
%
\section{Introduction}
\vspace{-0.2cm}
\label{sec:intro}
The CALICE collaboration~\cite{CALICE} is performing calorimeter development 
aiming to fulfill the hardware and physics demands of the International 
Linear Collider (ILC) physics program~\cite{ILC}. The ambitious required 
jet energy precision ($\approx 0.3/\sqrt{E(GeV)}$)
could be achieved with extremely segmented calorimeters using the particle 
flow approach~\cite{ILC}.

For the Analogue Hadron Calorimeter (AHCAL) option for the ILC a physics 
prototype was successfully used in test-beam campaigns at DESY, CERN 
and FNAL in the years \mbox{2006-09}, together with an electromagnetic 
calorimeter and a tail-catcher muon trigger prototype. 
That AHCAL prototype is a $38$ layer sampling calorimeter with 
$1$ m$^2$ lateral dimension, and total thickness of $4.5$ nuclear 
interaction lengths.
Each layer consists of \mbox{$2$ cm} thick steel absorber and a plane 
of $0.5$ cm thick plastic scintillator tiles. 
The tile sizes vary from $3$x$3$ cm$^2$ in the center of the layer, 
to $6$x$6$ cm$^2$ and $12$x$12$ cm$^2$ in the outer regions, 
resulting in $7608$ scintillating plates in total.
Each tile is coupled to a SiPM via a wavelength shifting fiber.
The physics prototype aims to verify the fulfilling of the requirements 
driven by the physics program forseen at the ILC using a 
highly granular and segmented tile hadronic calorimeter with 
silicon photomultiplier (SiPM) readout.
For the first time ever, SiPMs were used in a large scale 
real experiment, showing a stable behavior and efficiency.  
The main results on the detector calibration and data analysis
are reported in~\cite{CALICE_ANALYSIS}.

More recently, a new prototype for the AHCAL is under development 
and commissioning to demonstrate the feasibility to build a calorimeter
with fully integrated electronics meeting the constraints of a real 
detector. In order to cope with the expected channel number of the AHCAL
barrel of about 3.9 million, reliable assembly and production
procedures have to be developed. A new one-layer integration
prototype with about 2500 detector channels is currently under
development in order to test all assembly and integration
issues. 
%
\section{Technological Prototype Concept and Realization}
\begin{figure}[t!]
   \includegraphics[width=0.5\columnwidth, height=4.5cm]{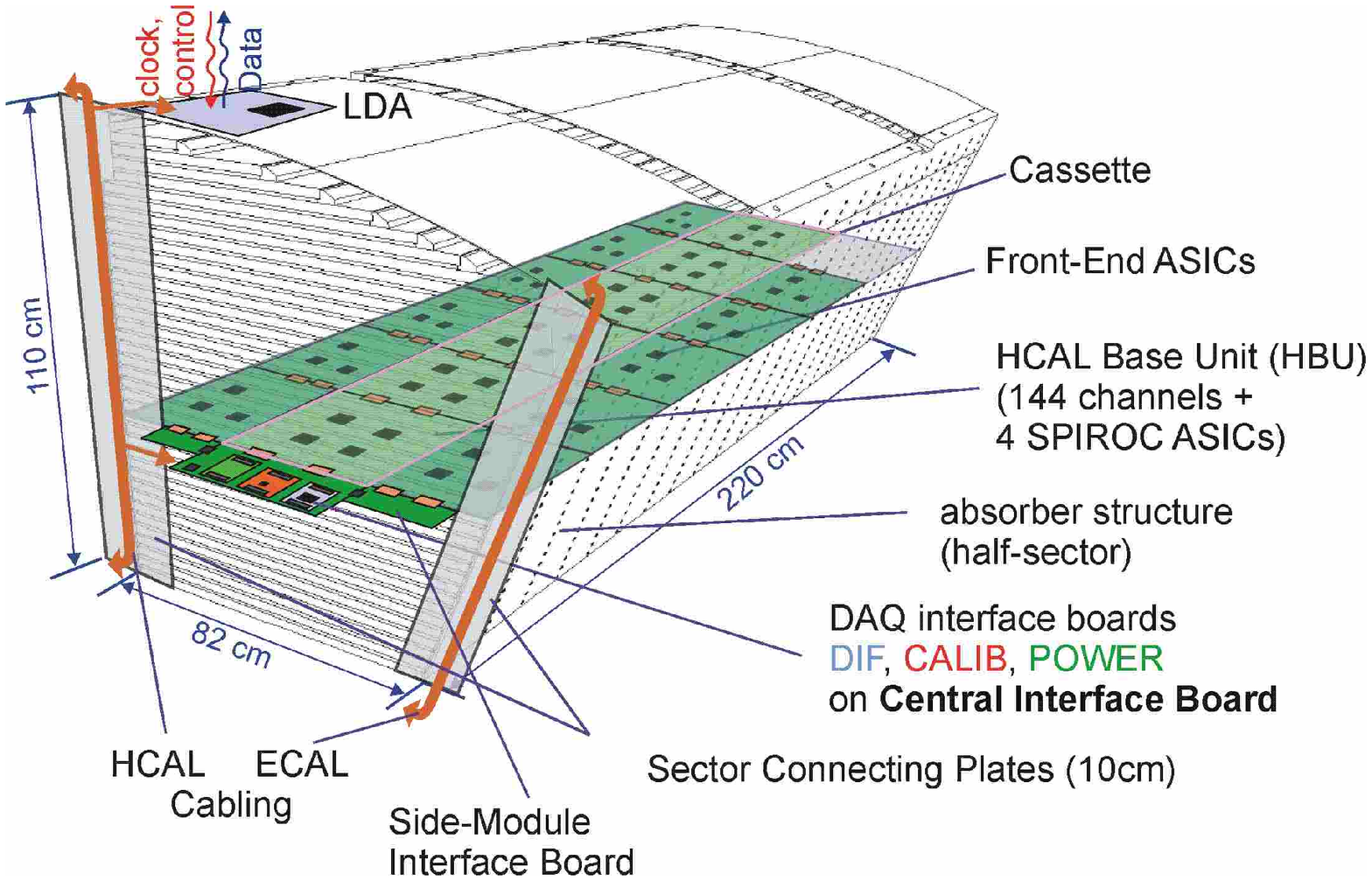}
   \hspace{0.5cm}
   \includegraphics[width=0.45\columnwidth, height=4.5cm]{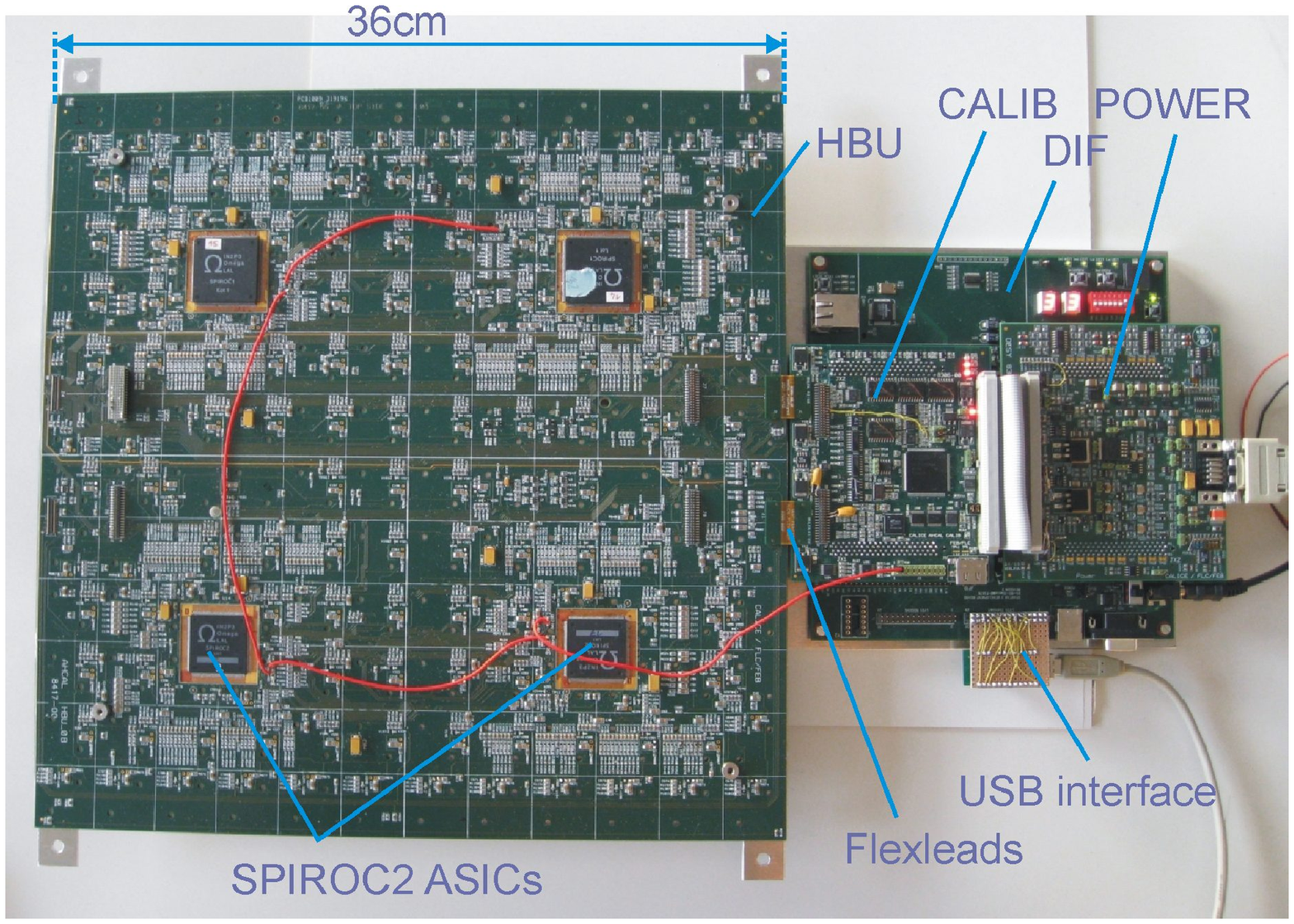}
   \vspace{-0.35cm}
   \caption{Left Panel: Drawing of half of an AHCAL section. 
            Right Panel: Setup of the HBU cassette at the 
            DESY test-beam area $22$. On the right of the cassette is 
            visible the CIB (DIF) module hosting the modules for the 
            SiPM calibration and for powering the system.}
   \label{fig:HBU}
   \vspace{-0.45cm}
\end{figure}

\vspace{-0.2cm}
\label{sec:prototype}
The circular structure of the AHCAL will be divided into
$16$ sectors. The resulting half
of a sector is shown in the left panel of Fig.~\ref{fig:HBU}. 

There are $48$ detector layers 
foreseen in the current proposal of a sector. Each layer consists 
of the tiles and the front-end electronics, integrated 
into the absorber structure and a $2$ cm thick stainless
steel absorber plate. The typical size of a layer is
\mbox{$1.0$x$2.2$ m$^2$}. 
The electronics of the sector layers is divided into
base units (HBUs) in order to keep the single modules at
reasonable sizes concerning the production and handling.
The HBU with a typical size of $36$x$36$ cm$^2$ integrates
144 scintillating tiles each with SiPMs together with the front-end
electronics and the light calibration system. The analogue
signals of the SiPMs are read out by four front-end low 
power dissipation SPIROC ASICs~\cite{SPIROC} designed at LAL 
Laboratory in Orsay. 
Each SPIROC handles the signal from $36$ independent channels.
After signal pre-amplification and shaping, the peaking amplitude
is stored sequentially in one of the $16$ cells (capacitors) of 
the analogue memory array, one array per channel.
In order to cover the full depth of \mbox{$220$ cm} of a sector layer, 
$6$ HBUs are connected together, forming an electrical layer unit 
within a layer cassette, a slab. 
Each sector layer will be equipped with three or four slabs.
The new integration prototype that is currently under development 
will cover one layer cassette.

The top view of the sofar realized prototype modules
is shown in the right panel of Fig.~\ref{fig:HBU}. 
Each module is equipped with two ASICs of the first SPIROC generation
and two of the current chip generation (SPIROC2). The results presented 
here refer to the two SPIROC2. 
On the right hand side of the HBU the interface
modules CALIB (for controlling the LED light-calibration
system, and for slow-control monitoring of temperature,
supply-voltages and -currents) and POWER (for providing 
all necessary inner detector supply voltages) are visible.  
They are connected as mezzanine modules to the Central 
Interface Board (CIB) module. 
The CIB is the only interface to the global data acquisition 
and hosts both, the fast interface path to the 
front-end ASICs and all slow-control tasks.
The size of the modules CALIB, POWER and CIB has to be
adapted to the final ILC mechanical environment in the next
integration step.
Details of the prototype realization can be found in~\cite{PROTOTYPES}.

%
\section{First Results}
\begin{figure}[t!]
   \begin{minipage}{7cm}
          \includegraphics[width=0.85\columnwidth, height=5.cm]
                          {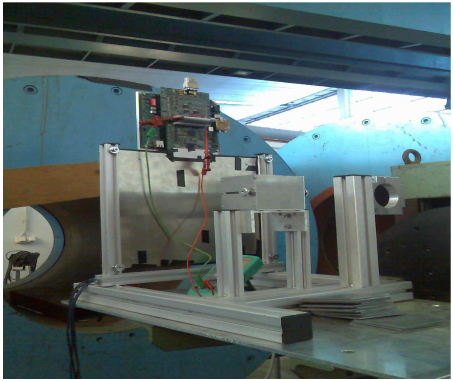}
   \end{minipage}
   \begin{minipage}{7.cm}
   \vspace{-0.9cm}
   \caption{The setup of the HBU cassette at the DESY test-beam area 
            $22$. On top of the cassette is visible the CIB module
            hosting the modules for the SiPM calibration and for 
            powering the system.}
   \label{fig:TESTBEAM}
   \end{minipage}
   \vspace{-0.65cm}
\end{figure}

\vspace{-0.2cm}
\label{sec:Results}
The commissioning of one module prototype is on going since 
this Spring at the DESY test-beam area, Fig.~\ref{fig:TESTBEAM}, while 
a second module is in the laboratory for debugging purposes.
In the laboratory environment, measurements 
help to understand all newly developed system components
and the system behavior. The full operation chain of the
system, including the programming of slow-control data, data
taking with external trigger or ASIC internal trigger, and data
readout is established. 

\begin{figure}[b!]
   \includegraphics[width=0.5\columnwidth, height=3.3cm]
                          {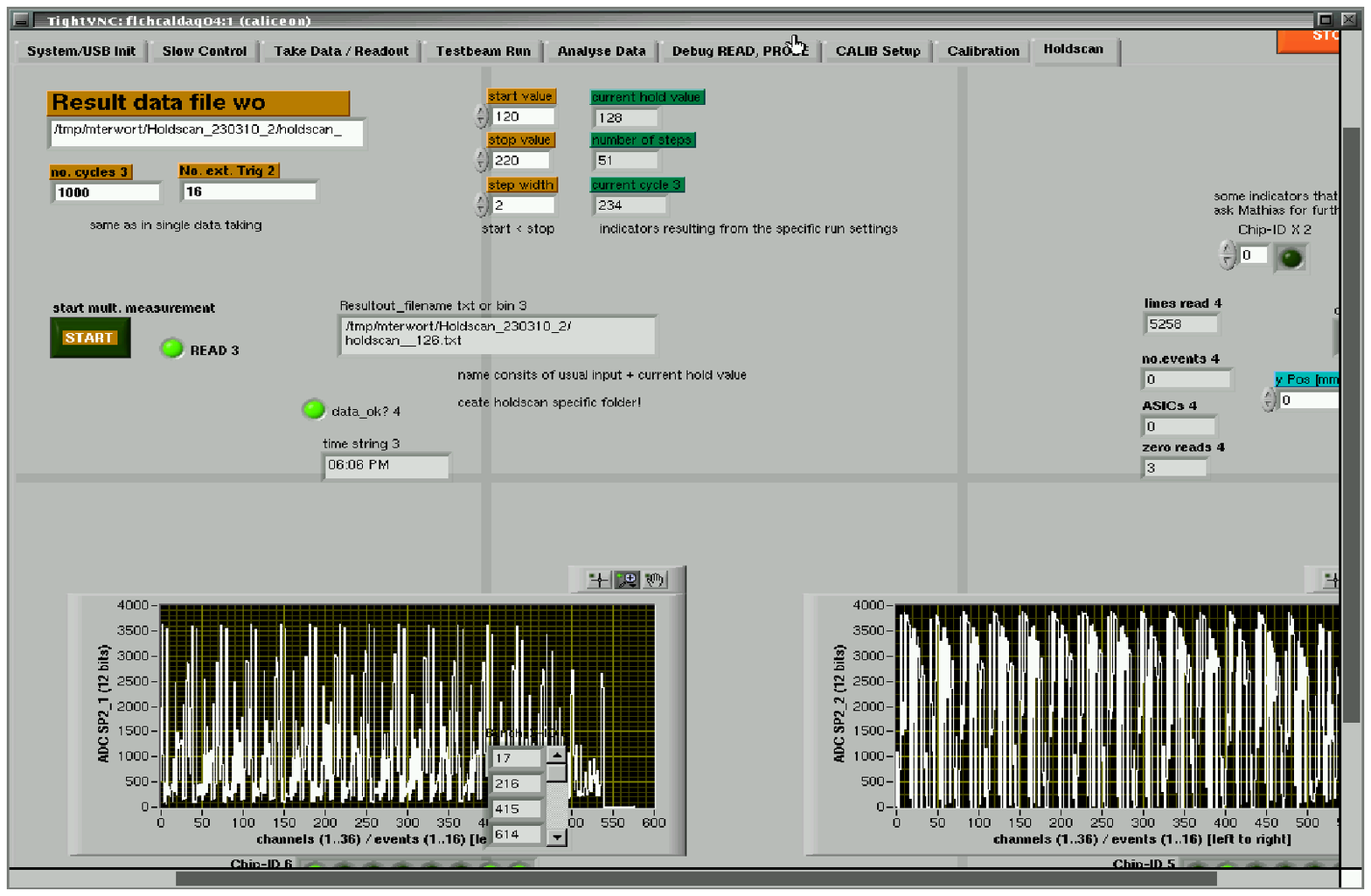}
   \hspace{0.5cm}
   \includegraphics[width=0.45\columnwidth, height=6.cm]
                          {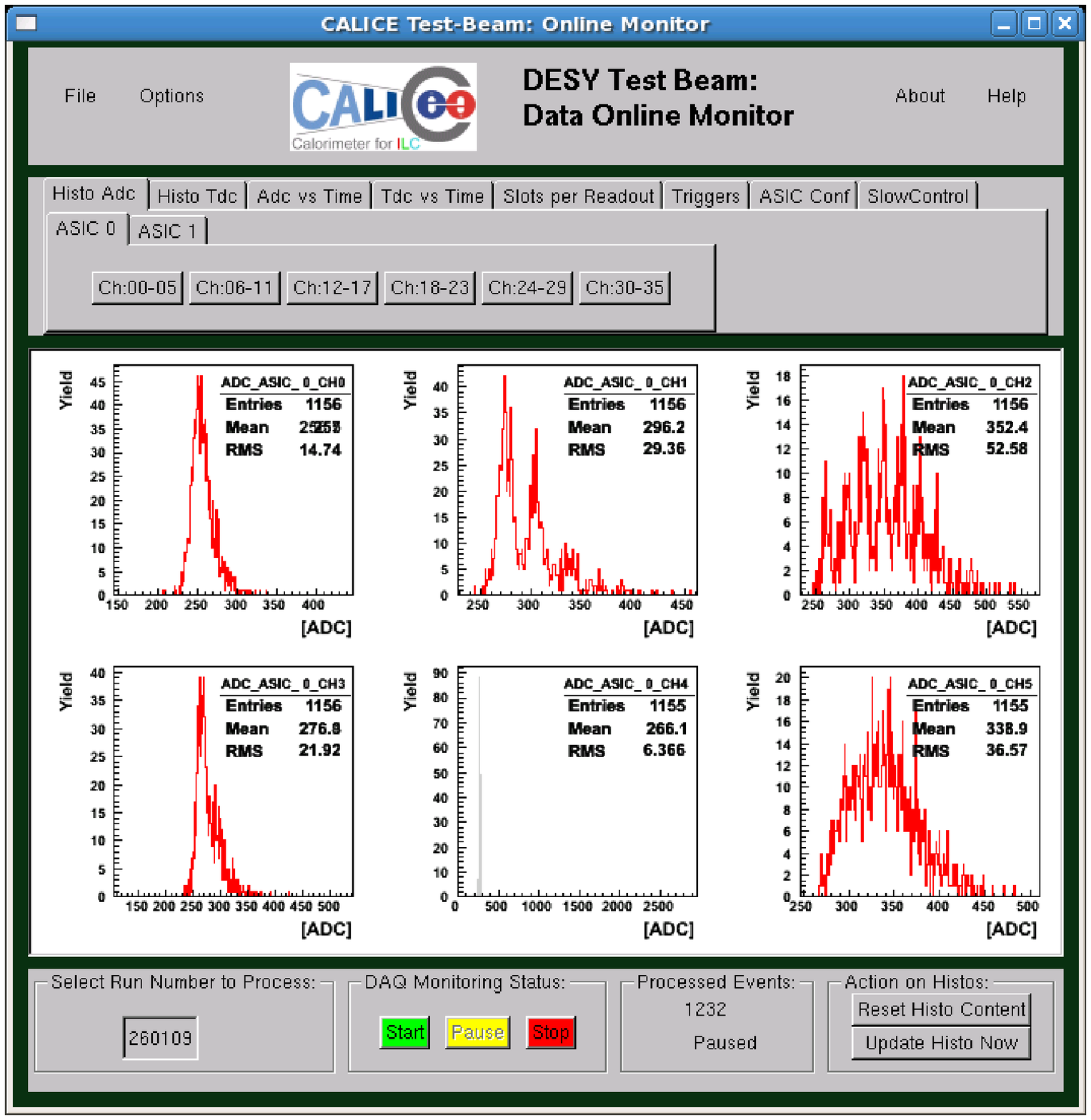}
   \vspace{-0.5cm}
   \caption{Screenshots of the DAQ program 
           (left panel) and of the online monitor GUI (right panel)
           in the test-beam control room taken during commissioning.}
   \label{fig:SOFTWARE}
\end{figure}

In the test-beam area the prototype is connected to the preliminary DAQ 
system that connects the system via USB-bus to a Linux PC with 
LabVIEW as graphical user interface, left panel of Fig.~\ref{fig:SOFTWARE}. 
When the prototype will be extended to the 
aimed 2500 detector channels (a three-slab layer), the DAQ will
be replaced by the final CALICE DAQ. Due to the data structure 
different from the previous CALICE test-beam campaigns, a new online monitor
has been written from scratch, scalable with the number of operating 
SPIROCs in the system, right panel of Fig.~\ref{fig:SOFTWARE}. 

\begin{figure}[t!]
   \includegraphics[width=0.5\columnwidth, height=5.5cm]
                          {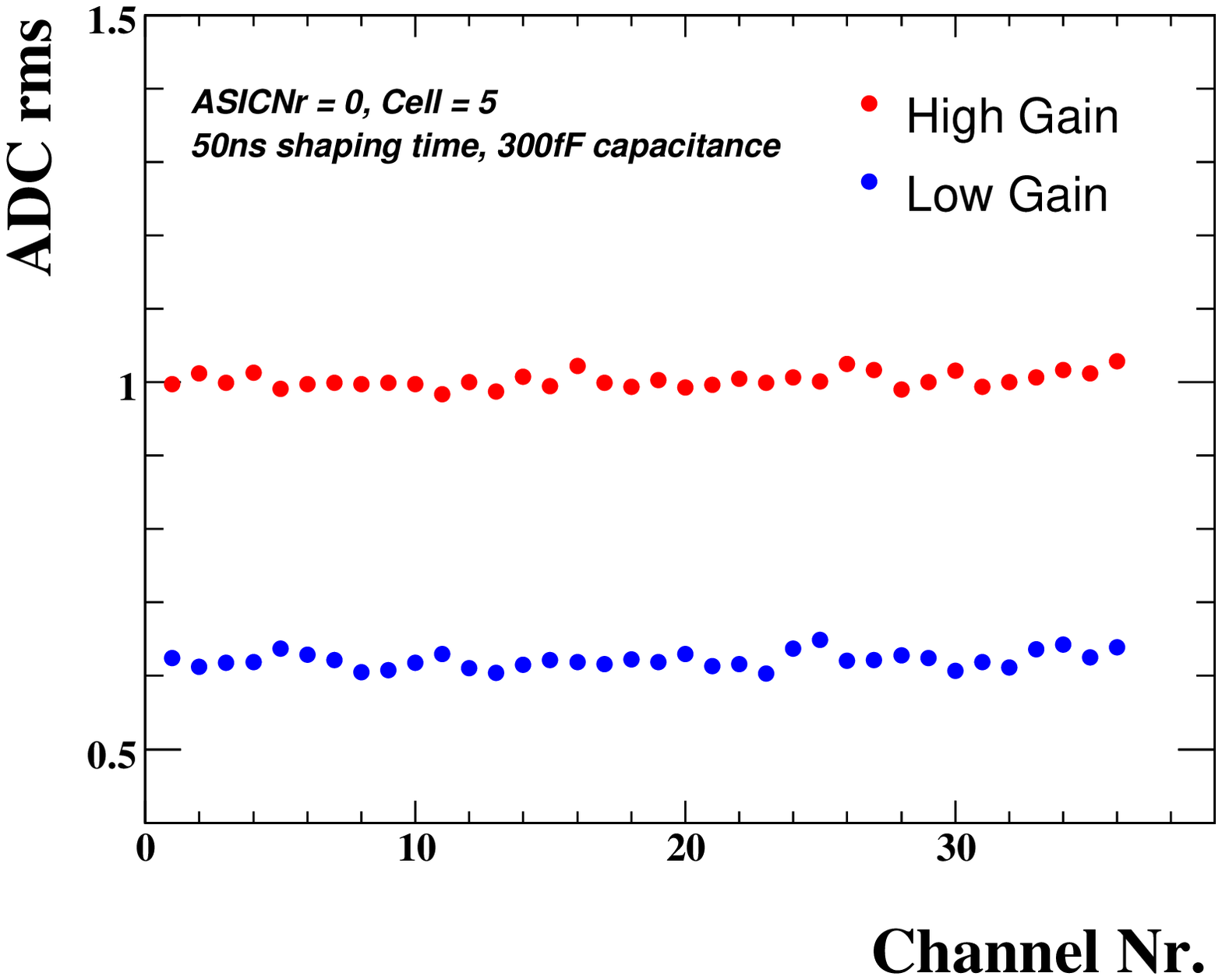}
   \includegraphics[width=0.5\columnwidth, height=5.cm]
                          {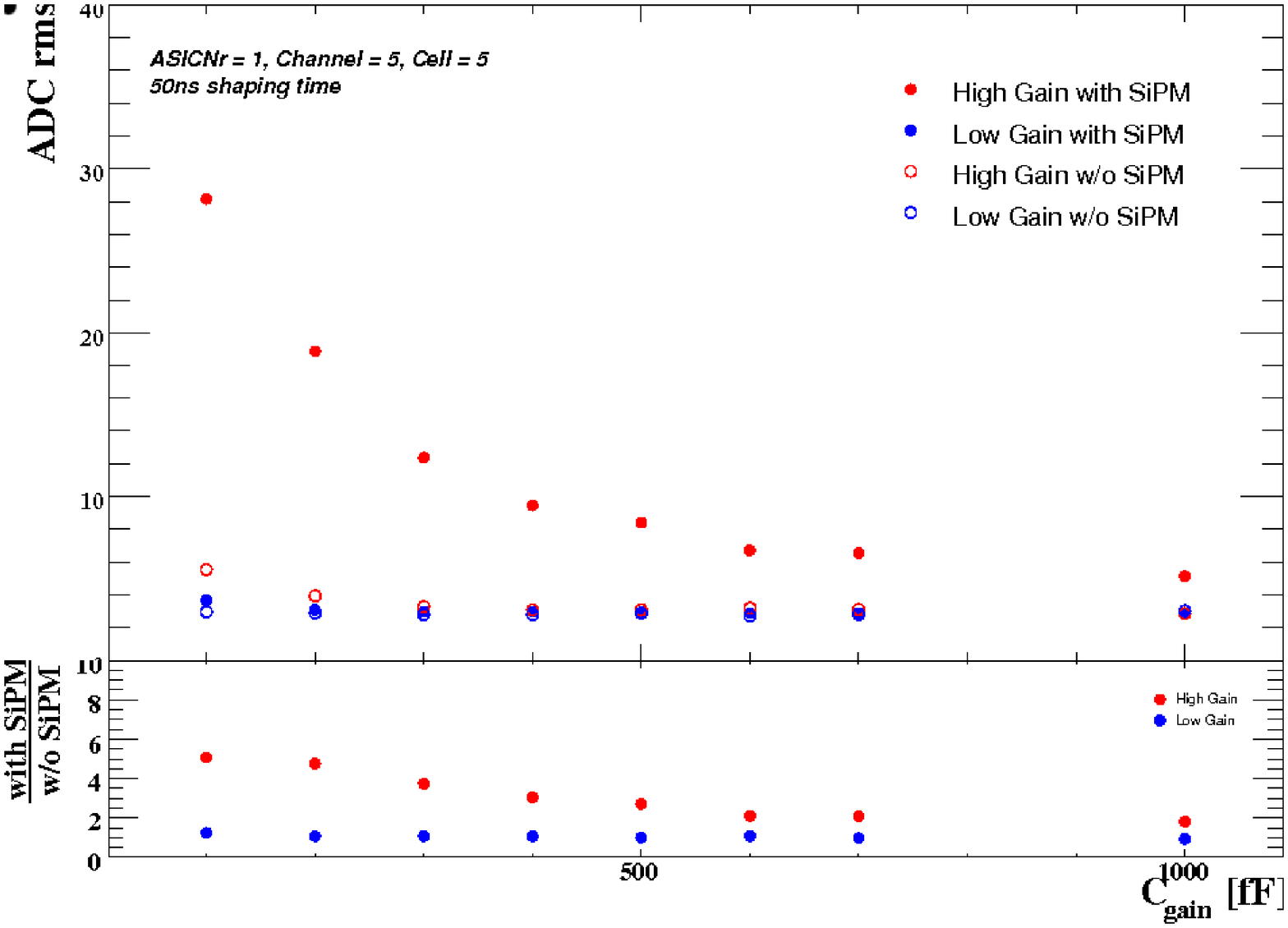}
   \vspace{-0.5cm}
   \caption{Example of noise measurements. 
            Left Panel: Uniformity between the $36$ channels of the ASICs.
            Right Panel: The noise is measured in high and low gain mode 
            for different values of the pre-amplification capacitance, 
            with and without powering on the SiPM.}
   \label{fig:noise}
   \vspace{-0.35cm}
\end{figure}

The first measurements were focused to the investigation of the 
noise level in all the channels for different setting
of the ASICs. An example of noise uniformity measurements is presented in 
the left panel of Fig.~\ref{fig:noise}.
The results are shown in terms of ADC units (two ADC units correspond
to $0.68$ mV, and one photoelectron peak corresponds to around 
$4\cdot 10^5$e operating the SiPMs at their nominal voltage). 
While the noise is sufficiently small and uniform for the $36$
channels of the presented ASIC, the second SPIROC shows a 
much higher and non-uniform noise, whose source is under investigation.
The noise was measured in high and low gain mode for different values 
of the pre-amplification capacitance, with and without powering on 
the SiPM.
An example of such measurements is presented in the right panel of 
Fig.~\ref{fig:noise}, showing that the system noise is 
dominated by the SiPMs.

\begin{figure}[t!]
   \includegraphics[width=0.5\columnwidth, height=4.5cm]
                          {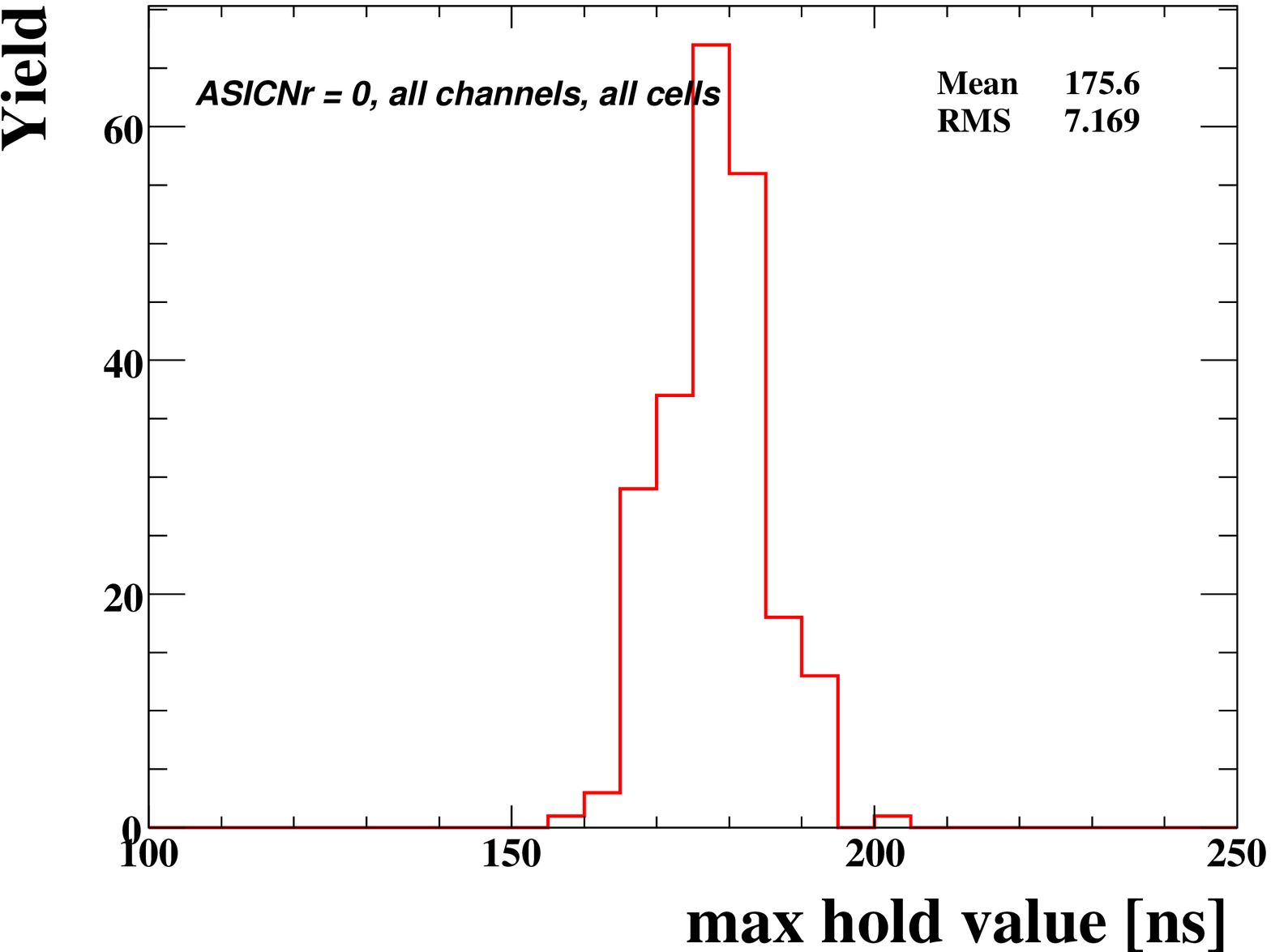}
   \includegraphics[width=0.5\columnwidth, height=5cm]
                          {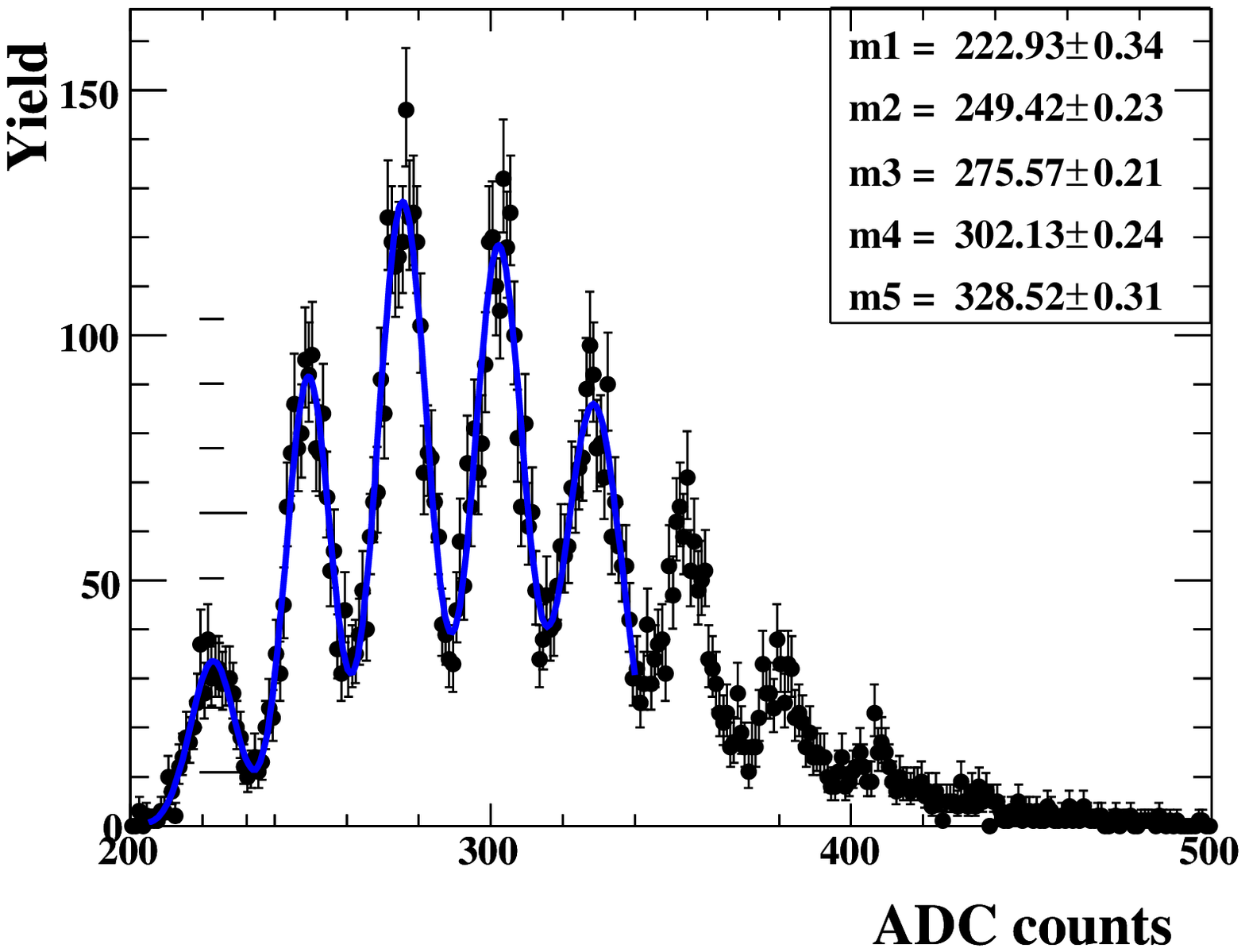}\\
   \vspace{-0.35cm}
   \begin{minipage}{8cm}
          \includegraphics[width=0.9\columnwidth, height=4.5cm]
                          {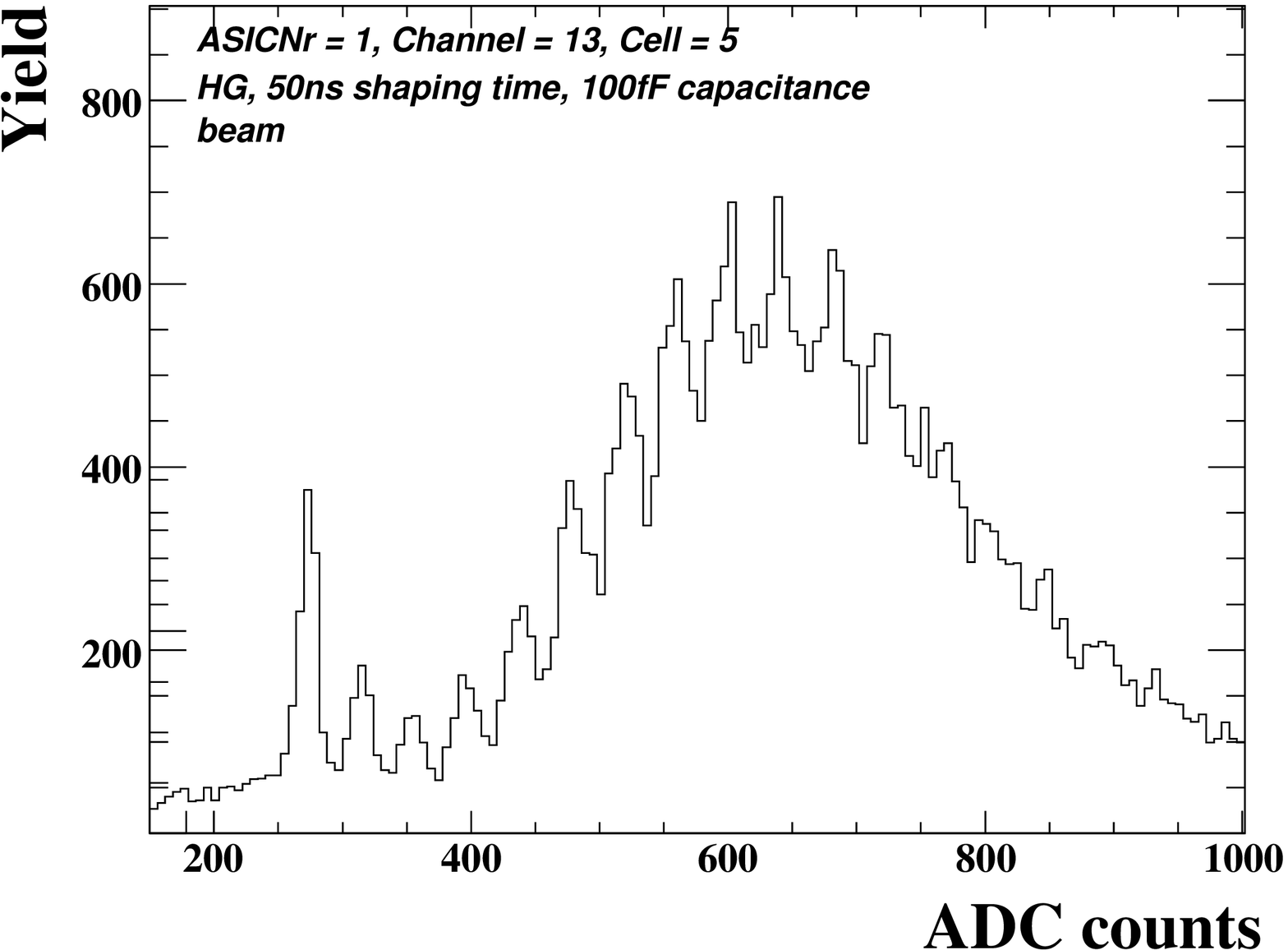}
   \end{minipage}
   \begin{minipage}{6.cm}
   \caption{Left top Panel: The distribution of the peaking time values 
            measured 
            for all channels and slots in the memory arrays of one ASIC.
            Right top Panel: The single-pixel spectra are fitted 
            with a multi-gaussian function to extract the SiPM gain.
            The extracted mean values of the gaussian peaks are 
            reported in the inserted panel.
            Bottom Panel: ADC spectrum for a mip-like particle.}
   \label{fig:hold}
   \end{minipage}
   \vspace{-0.15cm}
\end{figure}

By varying the delay of event triggers, signal values are eventually 
held inside the chip at different values of their amplitude, and the 
corresponding charge is then stored in the analogue memory. It is quite 
crucial to identify the proper trigger delay to store the value of the 
signal at its peaking amplitude. 
A scan of delay values was performed flashing the tiles with LED light 
of the calibration system~\cite{PROTOTYPES} to investigate the uniformity
of the peaking time for all the $16$ slots of all channels. 
The distribution of the measured peaking time values is shown for one 
ASIC in the top left panel of Fig.~\ref{fig:hold}. Here the scan was 
performed with \mbox{$5$ ns} steps. SPIROCs are operated at a 
fixed peaking time, and the observed $7$ ns spread was found 
(preliminary) to bias 
the measurement of the peaking time typically of less than $1\%$.
Similar measurements using mip-like particles (electrons) from 
the test-beam are on going. 

The gain of SiPMs was investigated flashing the tiles with 
low intensity LED light
and fitting the single-pixel ADC spectra with a multi-gaussian 
function, to eventually extract the distance between the peak 
maxima. An example of the procedure is presented in the top right 
panel of Fig.~\ref{fig:hold}.
The spread of the measured gain of the available SiPMs was found 
consistent with what reported in the datasheet provided by the manufactor.

A necessary step for establishing the electromagnetic energy scale 
of the prototype is the detection of mips. An example 
of mip spectrum taken using $2.5$ GeV test-beam electrons is shown 
in the bottom panel of Fig.~\ref{fig:hold}, where the ADC 
distribution has a maximum around $9$-$10$ pixels, in agreement 
with the datasheet parameters of SiPMs.

In summary, the technological prototype (with integrated electronics) 
of the AHCAL for the ILC is under development and commissioning 
to test all assembly and integration issues. The initial stage of 
the test-beam campaign at DESY has already helped in tuning the DAQ 
for commissioning the available modules. Several features of
the system have been investigated, as the noise level and its uniformity 
among the several channels, the peaking time determination  
for the signal from SiPMs, and the measurement of the gain of SiPMs.
Also, first mip-signals were observed, necessary for the determination 
of the electromagnetic scale of the detector. At the same time, 
close collaboration is on going with the ASIC designers to understand
the observed anomalous features of the SPIROC, in order to fix them 
in the next chip versions, possibly available for the next integration 
steps.

%
\section{Acknowledgments}
\vspace{-0.2cm}
 The author gratefully acknowledges P.\ G\"ottlicher, M.\ Reinecke,
 M.\ Terwort at DESY, and J.\ Sauer at the University of Wuppertal 
 for their valuable contribution to the results here presented.

%
\begin{footnotesize}

\end{footnotesize}

\end{document}